\documentstyle[aps,epsf]{revtex}  

\begin{document}        

\baselineskip 14pt
\title{Solar Neutrinos with Super-Kamiokande}
\author{Michael B Smy}
\address{University of California, Irvine}
%
\maketitle              

\begin{abstract}        
The discrepancy of the measured solar neutrino flux compared to the
predictions of the standard solar model may be explained by the neutrino
flavor oscillation hypothesis. A more direct and less model-dependent
test of this hypothesis is a measurement of the distortion of the shape
of the solar neutrino energy spectrum. Super-Kamiokande studies the
energy spectrum of recoil electrons from solar neutrino scattering
in water above 5.5 MeV.
\
\end{abstract}   	

\section{Introduction}               

Super-Kamiokande is a cylindrical Cherenkov detector containing
50,000 tons of purified water within a stainless steel tank.
The cylinder has a diameter of 39m and is 41m high.
It is optically divided into an inner detector (ID)
and an outer detector (OD). The OD is used as an
active shield of thickness $>2.75$m. The ID's diameter
is 34m and its height is 36m. From its 32,000 tons of water,
the inner 22,500 tons are used as fiducial volume. The fiducial
volume begins 2m from the ID surface. 11,146 inward-facing
50cm-diameter photomultiplier tubes (PMTs) cover $40 \%$ of
the light barrier between ID and OD. 1,885 outward-facing
20cm-diameter PMTs  view the OD.   

Super-Kamiokande is a Japanese-American experiment located 
about 300 kilometers northwest of Tokyo in the Japanese
Alps close to the village of Mozumi. To shield it from cosmic
rays, it was constructed underground in the Kamioka mine. The rock
overburden is 1000m (2,700m water equivalent). Next to a search
for nucleon decay, Super-Kamiokande was built to study solar
and atmospheric neutrinos as well as neutrinos originating from
galactic supernovae.

\section{Super-Kamiokande's Measurement of Solar Neutrinos}

Solar neutrinos are observed in Super-Kamiokande via elastic
neutrino-electron scattering in water:
\[
\nu_e+e^-\longrightarrow\nu_e+e^-
\]
The recoiling electron generates Cherenkov light which is detected
by the ID PMTs. The Cherenkov light pattern and intensity allow
reconstruction of the recoil electron's production time, production
location (vertex), direction and energy.

\begin{figure}[th]	

\noindent\epsfxsize 3.5 truein
\epsfbox{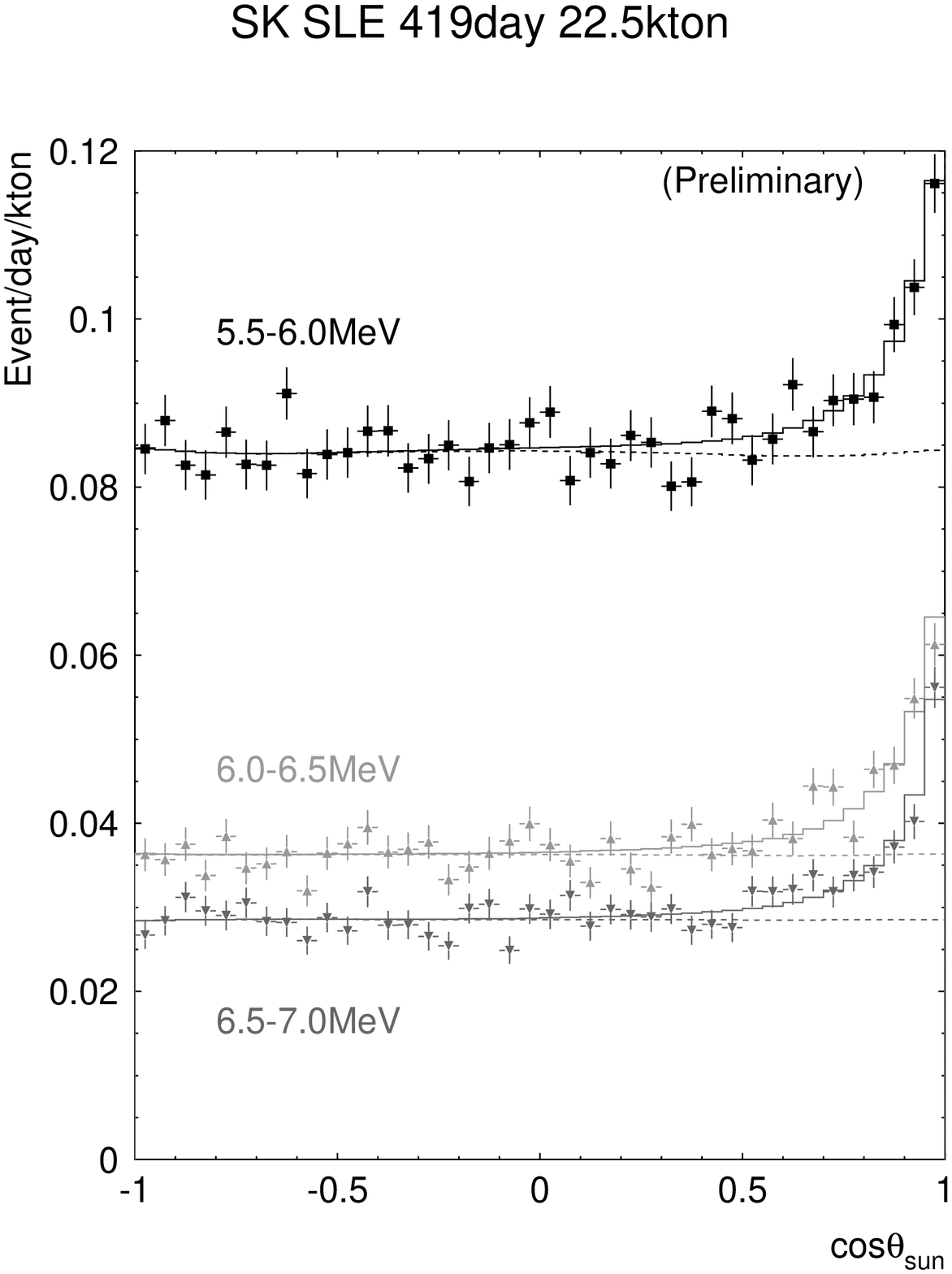}

\vskip -9cm\hspace*{3.4in}\epsfxsize 3.5 truein 
\epsfbox{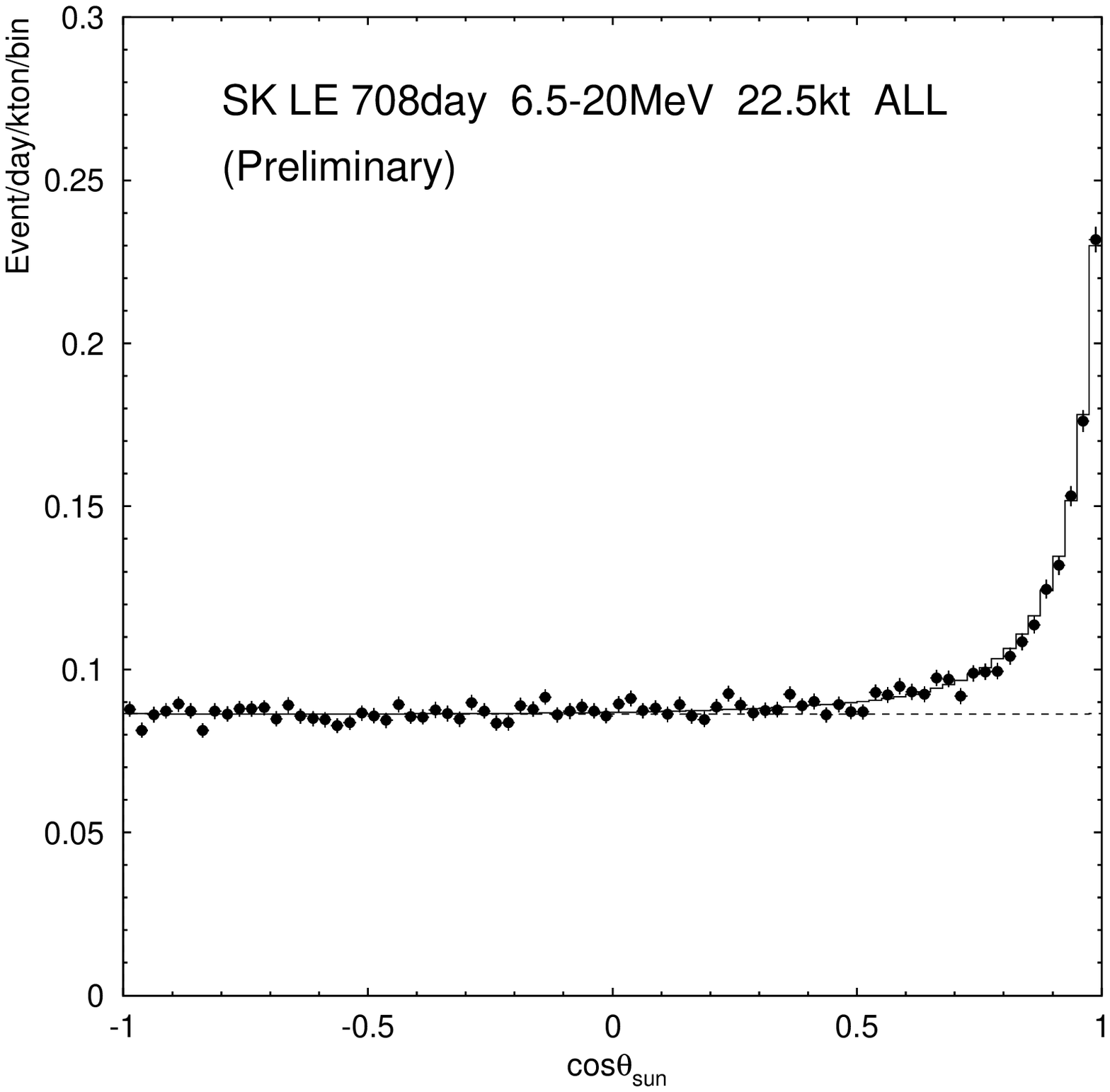}
\vskip 1cm

\caption[]{
\label{solsig}
\small a) $\cos \theta_{\mbox{Sun}}$ distributions below and \hspace*{1.5in}
b) above 6.5 MeV\\
The events close to $\cos \theta_{\mbox{Sun}}=1$ point back
to the sun. This `solar peak' above the background is used
to measure the solar $^8$B neutrino flux and spectrum. The
background increases dramatically with decreasing energy.
Super-Kamiokande has two analysis: the Super Low Energy (SLE)
analysis has a lower threshold but also lower event selection
efficiency than the Low Energy (LE) analysis.
}
\end{figure}

The reconstructed direction agrees within about thirty degrees of
the neutrino direction. Therefore the angle $\theta_{\mbox{Sun}}$
between reconstructed direction and the line drawn between the
Sun's current position and the reconstructed vertex can be used
to separate solar neutrino interactions from background events
(see figure~\ref{solsig}).
Since the interaction time is reconstructed, day-night and seasonal
variation of the solar neutrino flux can be studied. The reconstruction
of the recoil electron energy provides a lower bound on the neutrino
energy, therefore Super-Kamiokande can look for spectral distortions
in the solar neutrino flux. The recoil electron spectrum is 
sharply falling with increasing energy since higher energy neutrinos
can produce lower energy recoil electrons (see figure~\ref{solspec}).

Super-Kamiokande is sensitive to energies above 5 MeV, so the solar
neutrinos detected by it are produced by the reaction:
\[
^8\mbox{B}\longrightarrow ^8\!\mbox{Be}^*+e^++\nu_e
\]
This reaction has an endpoint of 15 MeV (see figure~\ref{solspec}).
Super-Kamiokande obtained 
$9397^{+154}_{-144}$(stat.)$^{+263}_{-254}$(syst.) signal events
above 6.5 MeV recoil electron energy. From this number the total
$^8$B solar neutrino flux can be calculated to be
($2.44\pm0.04$(stat.)$\pm0.07$(syst.))$\times 10^6/$cm$^2$/sec. This
is lower than expected from the BP98\cite{bp98} Standard Solar Model (SSM).
The ratio is
$
\frac{\mbox{Data}}{\mbox{SSM$_{\mbox{BP98}}$}}=
0.471^{+0.008}_{-0.007}\mbox{(stat.)}^{+0.013}_{-0.013}\mbox{(syst.)}
$

\section{Solar Neutrino Spectrum}

The discrepancy between SSM and flux measurement (one aspect of the
so-called solar problem) may be explained
by neutrino flavor oscillation, where electron-type neutrinos
oscillate to other, undetected neutrino types and back. Indeed, if
they oscillate to just one other type, and if the oscillation length
is short compared to the distance sun---earth, a suppression of
about $50\%$ is not unexpected. This neutrino oscillation hypothesis
seems more likely now, since atmospheric muon-type neutrinos indeed
appear to oscillate to other types\cite{atmos}.

\begin{figure}[th]	

\noindent\epsfxsize 3.5 truein
\vspace*{2cm}

\epsfbox{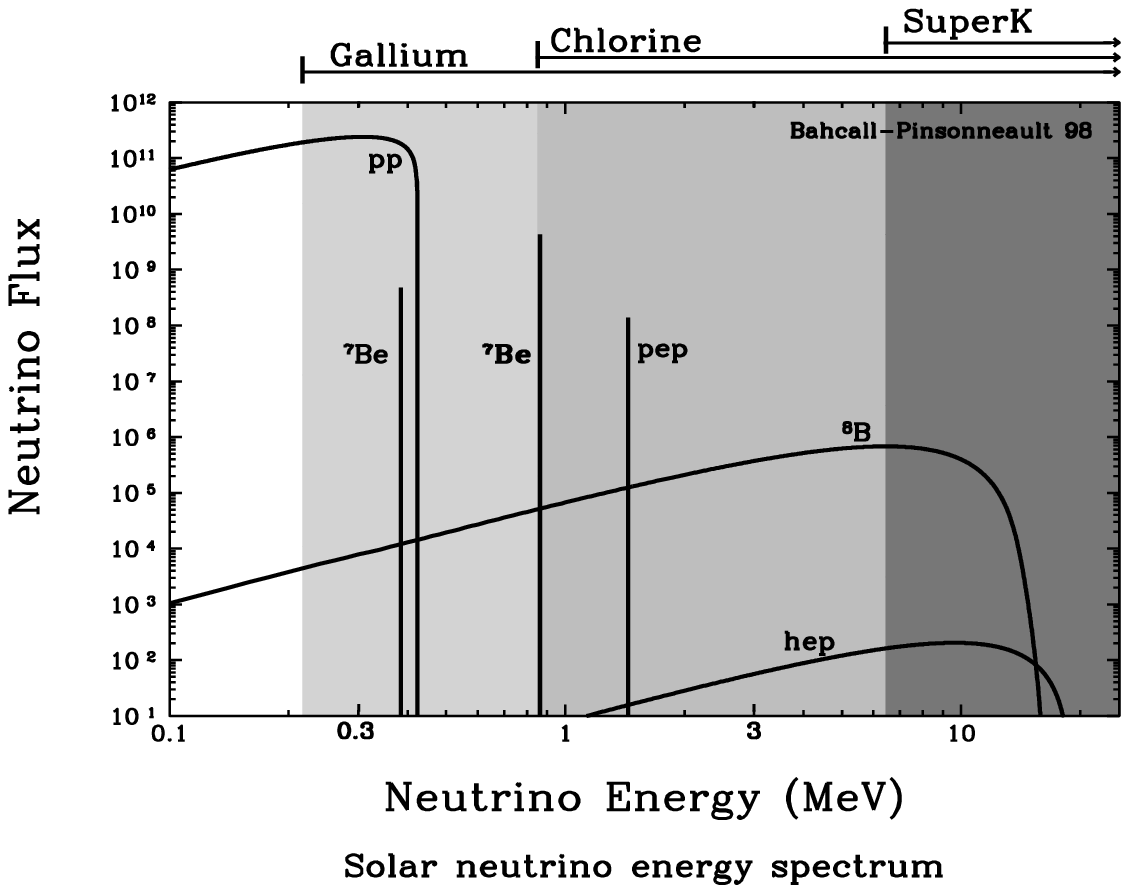}

\vskip -9cm\hspace*{3.4in}\epsfxsize 3.5 truein 
\epsfbox{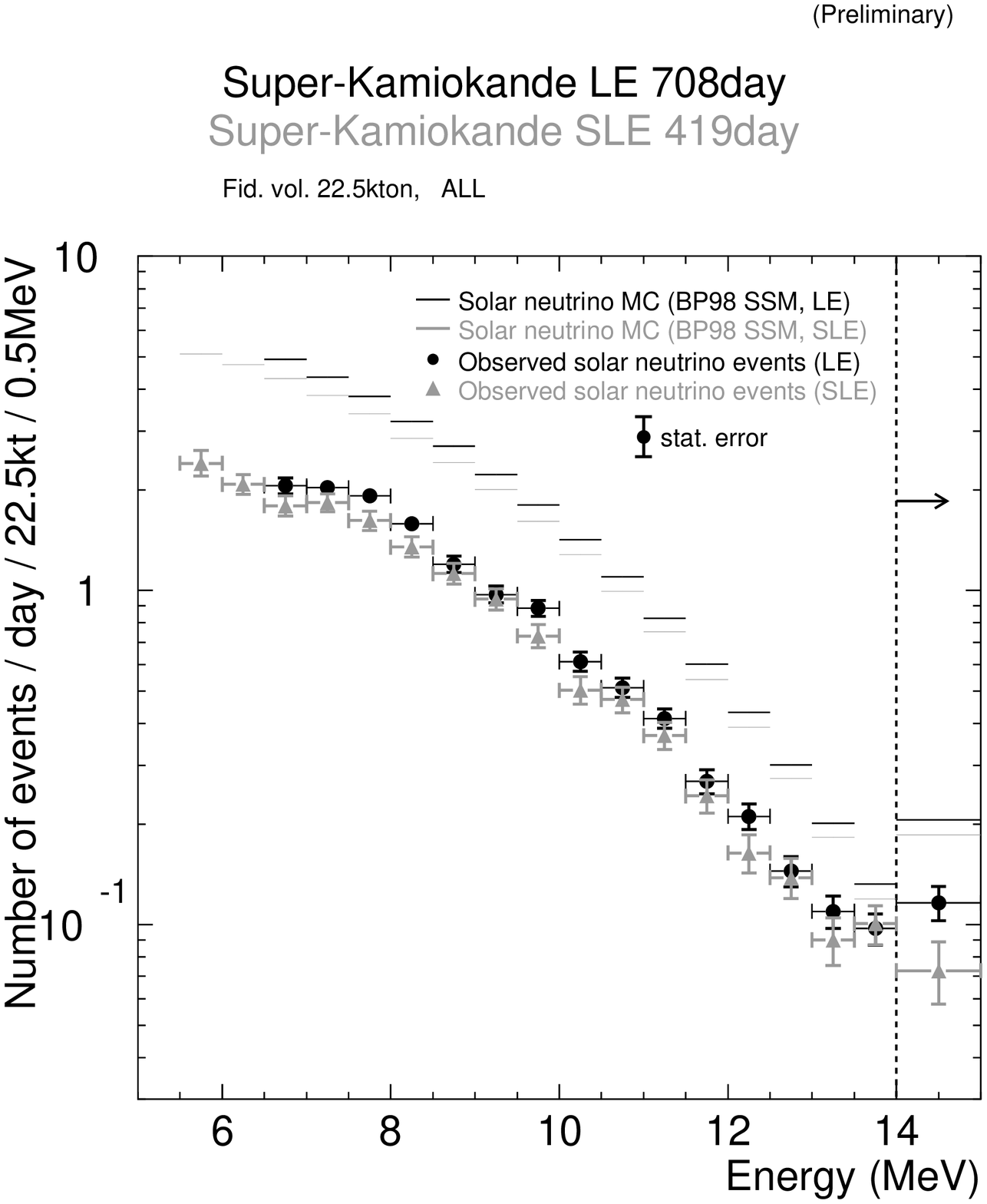}

\caption[]{
\label{solspec}
\small a) Solar Neutrino Energy Spectrum (BP98\cite{bp98})\hspace*{0.2in}
b) Measured Recoil Electron Spectrum from $^8$B Neutrinos\\
Super-Kamiokande's new Super Low Energy (SLE) analysis has a lower
threshold but a lower event selection efficiency.}
\end{figure}

However, a more model-independent way of testing this oscillation
hypothesis is to study the energy spectrum, the day--night and the
seasonal variation of the solar flux, rather than relying on flux
measurements alone. The SSM prediction of the $^8$B flux can be
seen in figure~\ref{solspec}. Super-Kamiokande has measured the
recoil electron spectrum\cite{spec} and the day--night
variation\cite{daynight} above 6.5 MeV using the Low Energy (LE)
analysis. Below 6.5 MeV the background of this analysis increases
sharply with decreasing energy. A new Super Low Energy (SLE) analysis
was designed to reject these backgrounds more efficiently.
Currently, the SLE analysis threshold is at 5.5 MeV.
The LE analysis has a higher event selection efficiency than the
Super Low Energy (SLE) analysis. Therefore both analysis are
combined to study the spectrum: above 6.5 MeV the LE analysis is used, and
below 6.5 MeV the SLE analysis.

A convenient way to search for spectral distortions is to normalize
the observed spectrum by the SSM expectations (see figure~\ref{normspec}~a)).
A flavor oscillation hypothesis leads to deviations from a flat normalized
spectrum. The strongest deviations are expected for ``vacuum oscillations''
at small $\Delta m^2$ (see figure~\ref{normspec}~b)). Matter-enhanced
(MSW) neutrino oscillations~\cite{msw} show smaller deviations. In particular,
the large-angle solution expects an almost flat normalized spectrum.
The small-angle solution leads to a sloped normalized spectrum (low
recoil energy bins are more strongly suppressed than high energy bins).
The best oscillation fit to Super-Kamiokande's spectrum is for
$\sin^2 2\theta=0.87$, $\Delta m^2=4.3\cdot 10^{-10}$ eV$^2$ in the
vacuum oscillation area.

\subsection{Hep Neutrinos}

\begin{figure}[th]	

\noindent\epsfxsize 3.5 truein
\epsfbox{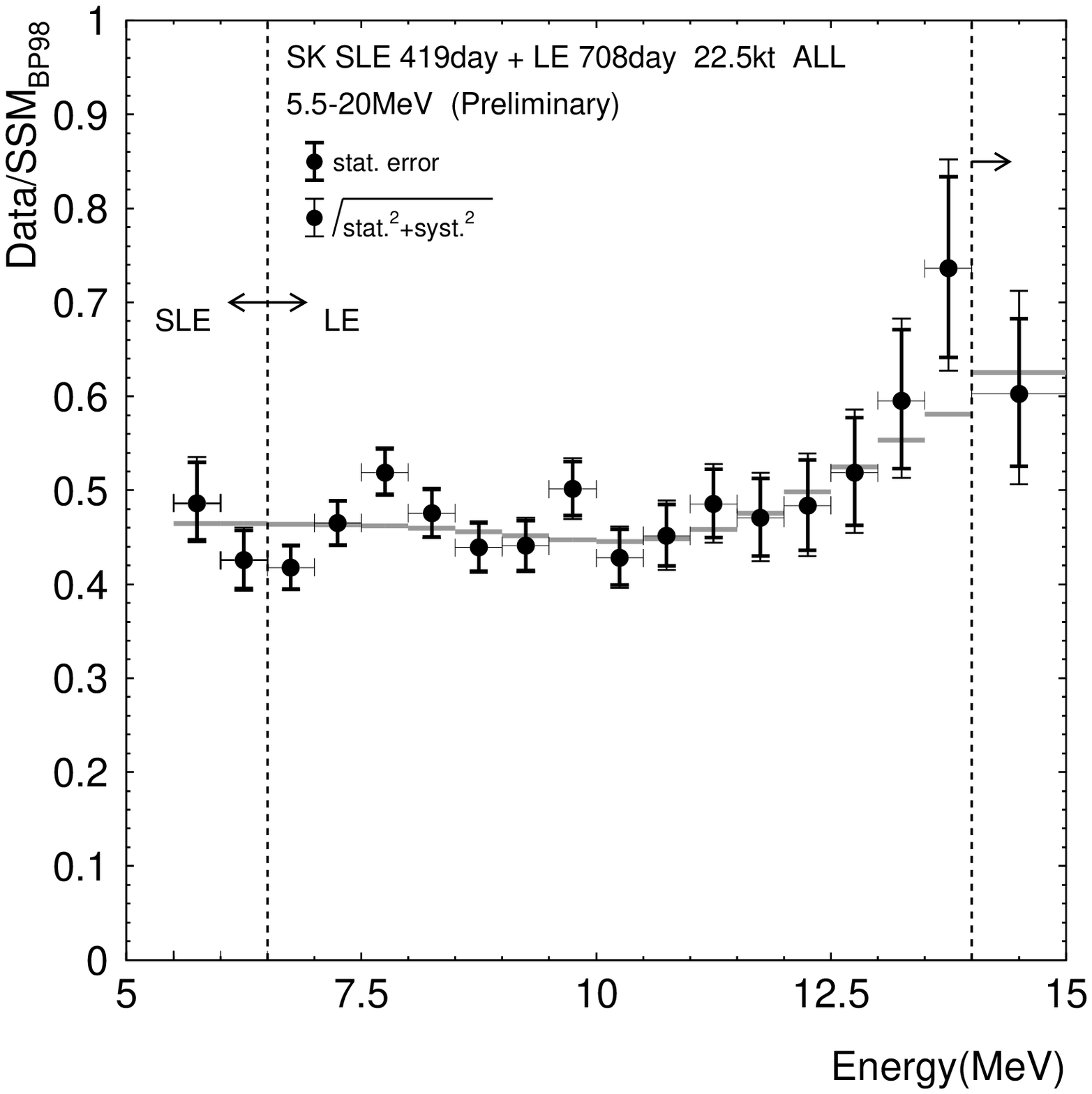}
\epsfxsize 3.5 truein 
\epsfbox{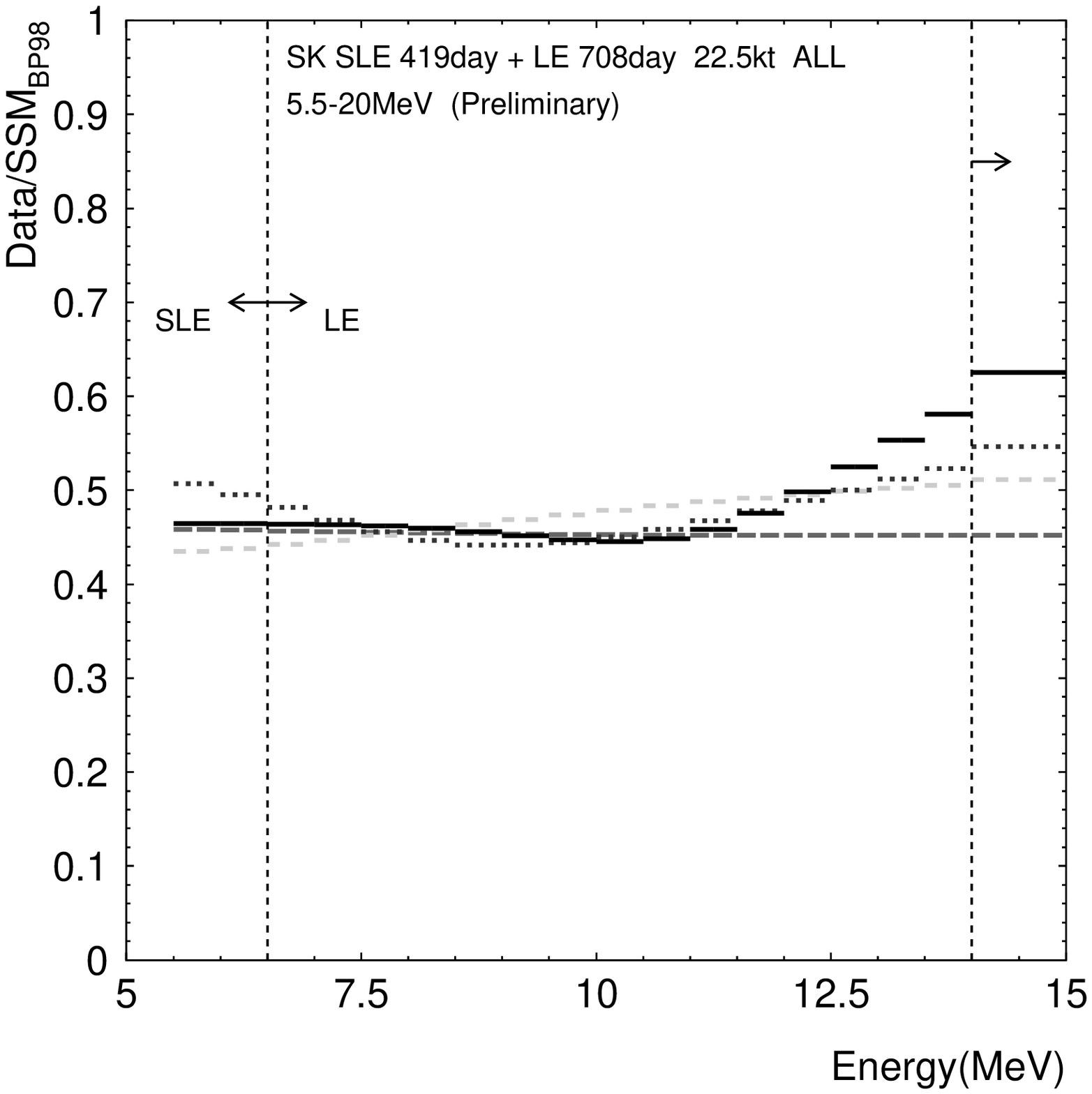}
\vspace*{0.3in}

\caption[]{
\label{normspec}
\small a) Normalized Recoil Electron Spectrum \hspace*{0.8in}
b) Expected Spectra for Different Oscillation Parameters\\
The best fit to the spectrum is vacuum oscillation with
$\sin^2 2\theta=0.87$, $\Delta m^2=4.3\cdot 10^{-10}$ eV$^2$
(grey line in a), solid black line in b)).
Small-angle ($\sin^2 2\theta=0.0045$,
$\Delta m^2=5.6\cdot 10^{-6}$ eV$^2$) and large-angle
($\sin^2 2\theta=0.7$, $\Delta m^2=2.8\cdot 10^{-5}$ eV$^2$)
MSW oscillation expectations can also be seen in b)
(light grey and dark grey dashed line). The dark grey, dotted
line is vacuum oscillation expectation with lower $\Delta m^2$
($\sin^2 2\theta=0.83$, $\Delta m^2=7.5\cdot 10^{-11}$ eV$^2$).}
\end{figure}

\begin{figure}[th]	

\noindent\epsfxsize 3.5 truein
\centerline{\epsfbox{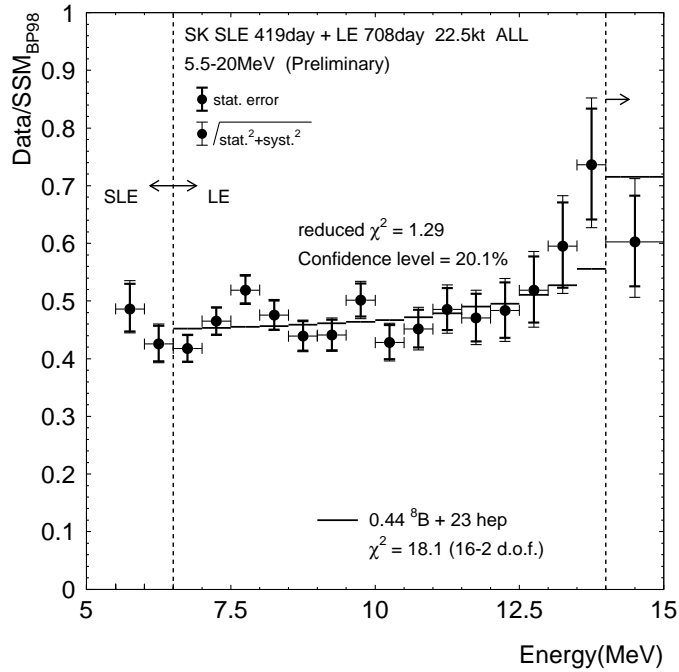}}
\vspace*{0.3in}

\caption[]{
\label{hepspec}
\small Best no oscillation fit without constraint on the
$^8$B or Hep flux normalization.}
\end{figure}

The deviations from a flat normalized spectrum in Super-Kamiokande
arise mainly at the high energy end of the spectrum. The rate of
solar neutrino emission from the process
\[
^3\mbox{He}+p\longrightarrow ^4\!\mbox{He}+e^++\nu_e
\]
(Hep flux) contributes only a tiny amount to
Super-Kamiokande's flux as expected by the SSM (see figure~\ref{solspec}).
However, the SSM Hep flux normalization is highly uncertain. An increase
in the Hep flux of a factor of about twenty times with respect to the SSM
will explain the deviations shown in figure~\ref{normspec} a) (see also
figure~\ref{hepspec}). However, the best oscillation fit to the data is still
in the vacuum oscillation region.

\subsection{Super Low Energy Analysis}

The normalized recoil electron spectrum at the low end of Super-Kamiokande's
sensitivity does not suffer from the uncertainty of the Hep flux. The MSW
small-angle area predicts a suppression of the flux with respect to the
flux around 10 MeV, for large-angle MSW oscillation a small increase is
expected. Vacuum oscillations of different $\Delta m^2$ expect different
behavior. 

To be sensitive to these low energy deviation, both new hardware and
software developments were necessary. Super-Kamiokande's normal
energy threshold of around 5.5 MeV leads to a fairly small and strongly
energy-dependent trigger efficiency in the SLE region. To increase
the efficiency and decrease the energy dependence, the trigger
threshold had to be lowered to around 4.5 MeV. In order to do that,
it is necessary to use an ``intelligent trigger'' to handle the
resulting high raw trigger rate. This intelligent
trigger, a fast computer for online filtering of very low
energy events, was installed in May 1997. It reduces the raw trigger rate
of about 100 Hz to about 5 Hz. New filtering software exploits the
observation, that background events seem to be concentrated close
to the wall of the ID.
Currently, the SLE analysis has a threshold of 5.5 MeV and is adding
two energy bins to the LE analysis. The 5-5.5 MeV energy bin will soon
be added to the analysis. The SLE analysis results above 6.5 MeV 
agree with the LE analysis.

\section{Day--Night Variation}

\begin{figure}[th]	

\noindent\epsfxsize 3.5 truein
\epsfbox{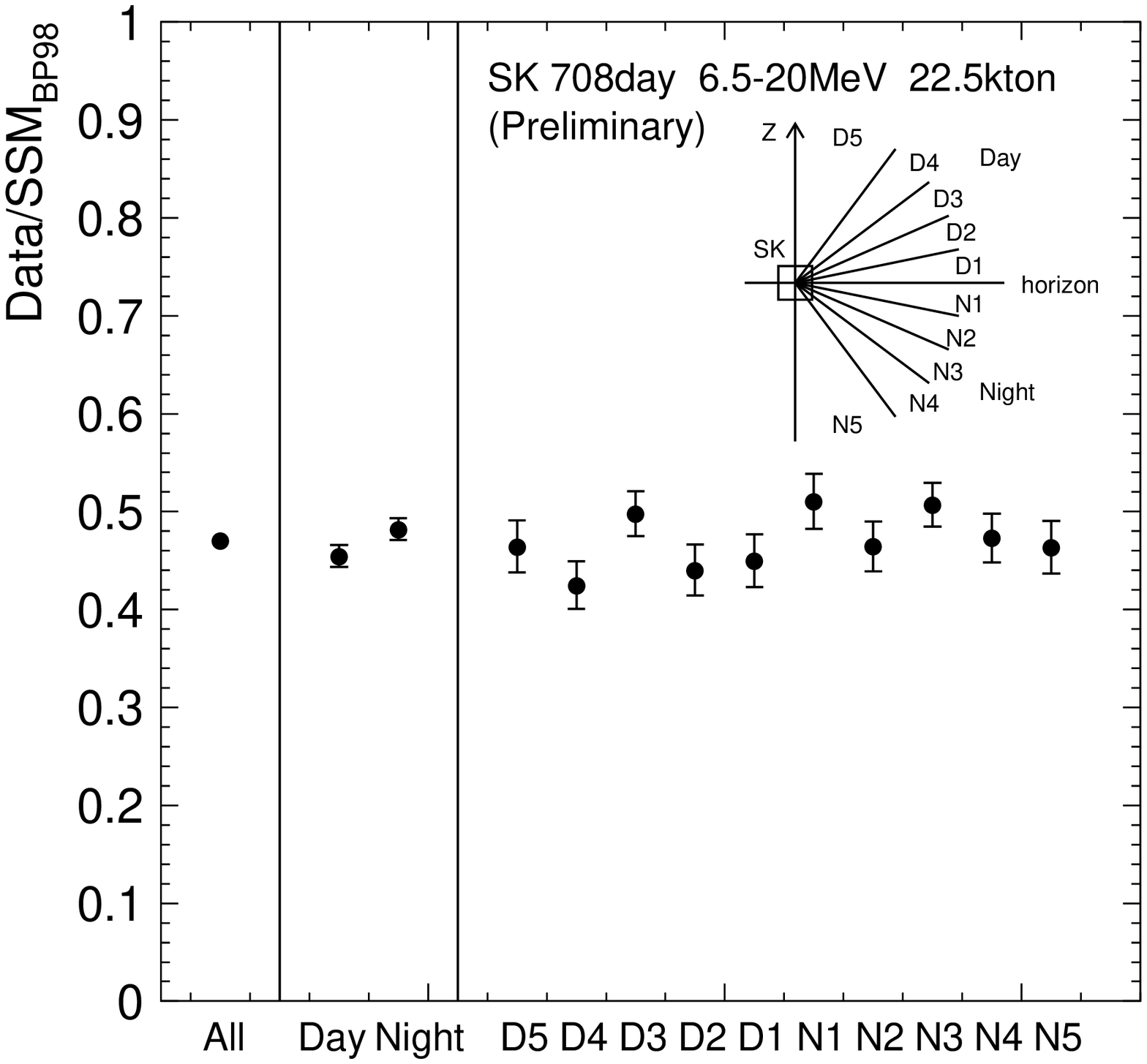}
\epsfxsize 3.5 truein 
\epsfbox{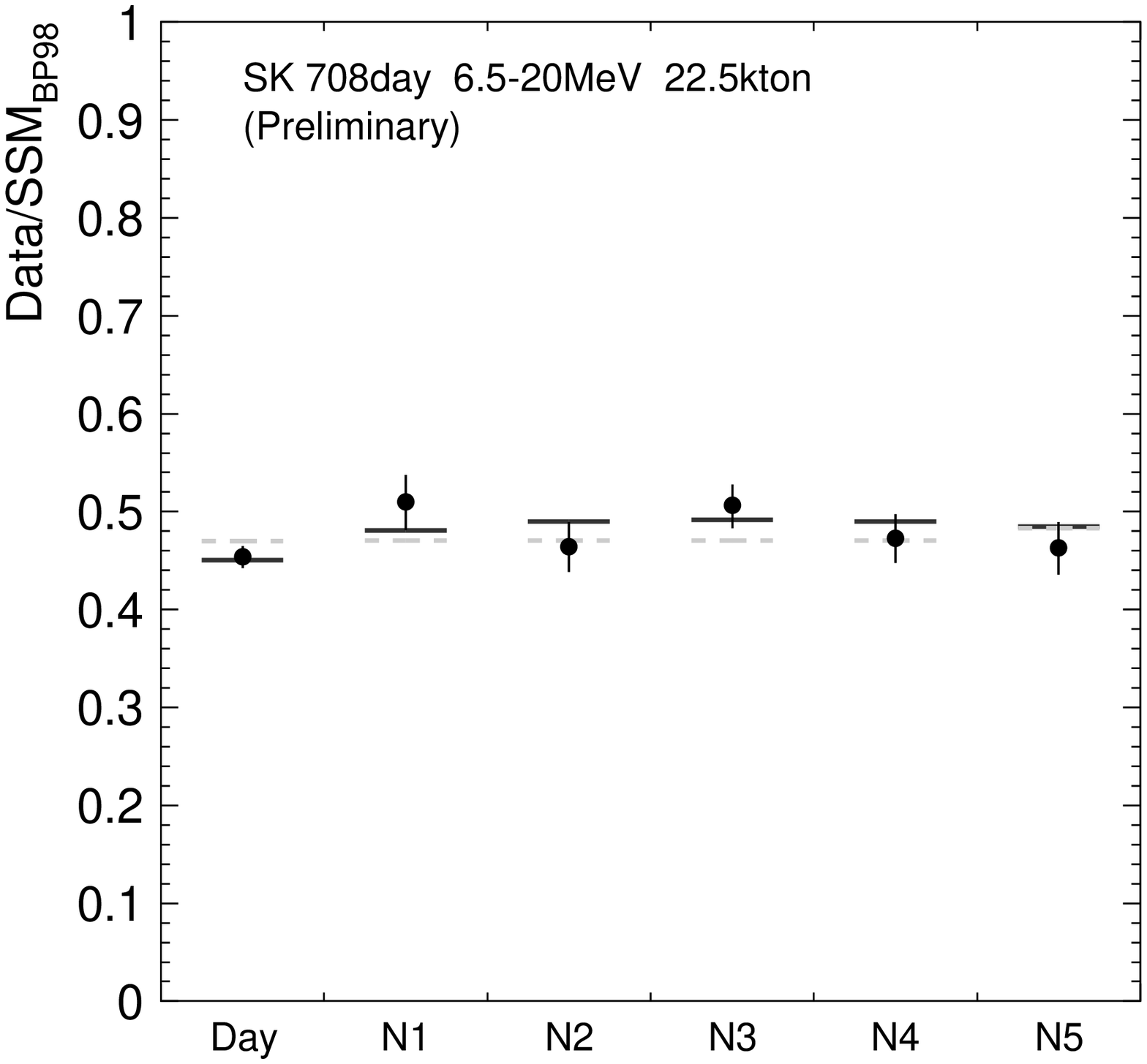}
\vspace*{0.2in}

\caption[]{
\label{daynight}
\small a) Day--Night Asymmetry and Variation \hspace*{0.2in}
b) Expected Day--Night Variation for Different Oscillation Parameters\\
The day-night asymmetry of the flux above 6.5 MeV displayed in a)
is $-0.026\pm0.016$(stat.)$\pm0.013$(syst.) Ten zenith angle bins
are defined as shown in a).
The best fit to the data is MSW large-angle oscillation with
$\sin^2 2\theta=1$, $\Delta m^2=1.9\cdot 10^{-5}$ eV$^2$
(solid dark grey line in b)). The small-angle MSW oscillation
expectation (dashed light grey line in b) with
$\sin^2 2\theta=0.005$, $\Delta m^2=7.1\cdot 10^{-6}$ eV$^2$)
shows a core enhancement (night 5 bin) which is disfavored by
the data.}
\end{figure}

MSW oscillations predict an enhancement of the solar neutrino flux during
the night when the neutrinos pass through the earth. Super-Kamiokande
measures the neutrino flux above 6.5 MeV as a function of the zenith
angle, the angle of the line sun --- Super-Kamiokande with respect to
the vertical (see figure~\ref{daynight}). The small-angle solution
expects only an increase in flux when the neutrinos pass through the core of
the earth. This effect would only appear in the `night 5' bin in
figure~\ref{daynight} b) which corresponds to a zenith angle of
about $140^0$ to $180^0$. The large-angle
solution predicts an enhancement throughout the night. The best
oscillation fit to Super-Kamiokande's zenith angle variation is 
$\sin^2 2\theta=1$, $\Delta m^2=1.9\cdot 10^{-5}$ eV$^2$.

The measured day--night asymmetry defined as
\[
\frac{\mbox{day}-\mbox{night}}{\mbox{day}+\mbox{night}}=
-0.026\pm0.016\mbox{(stat.)}\pm0.013\mbox{(syst.)}
\]
differs not significantly from zero. However, even the largest expected
asymmetry is still consistent with the data. The data disfavors the
hypothesis of a `core effect', but slightly favors a negative day--night
asymmetry. More data needs to be taken to resolve the predicted matter effects.

\section{Seasonal Variation}

\begin{figure}[th]	

\noindent\epsfxsize 3.5 truein
\epsfbox{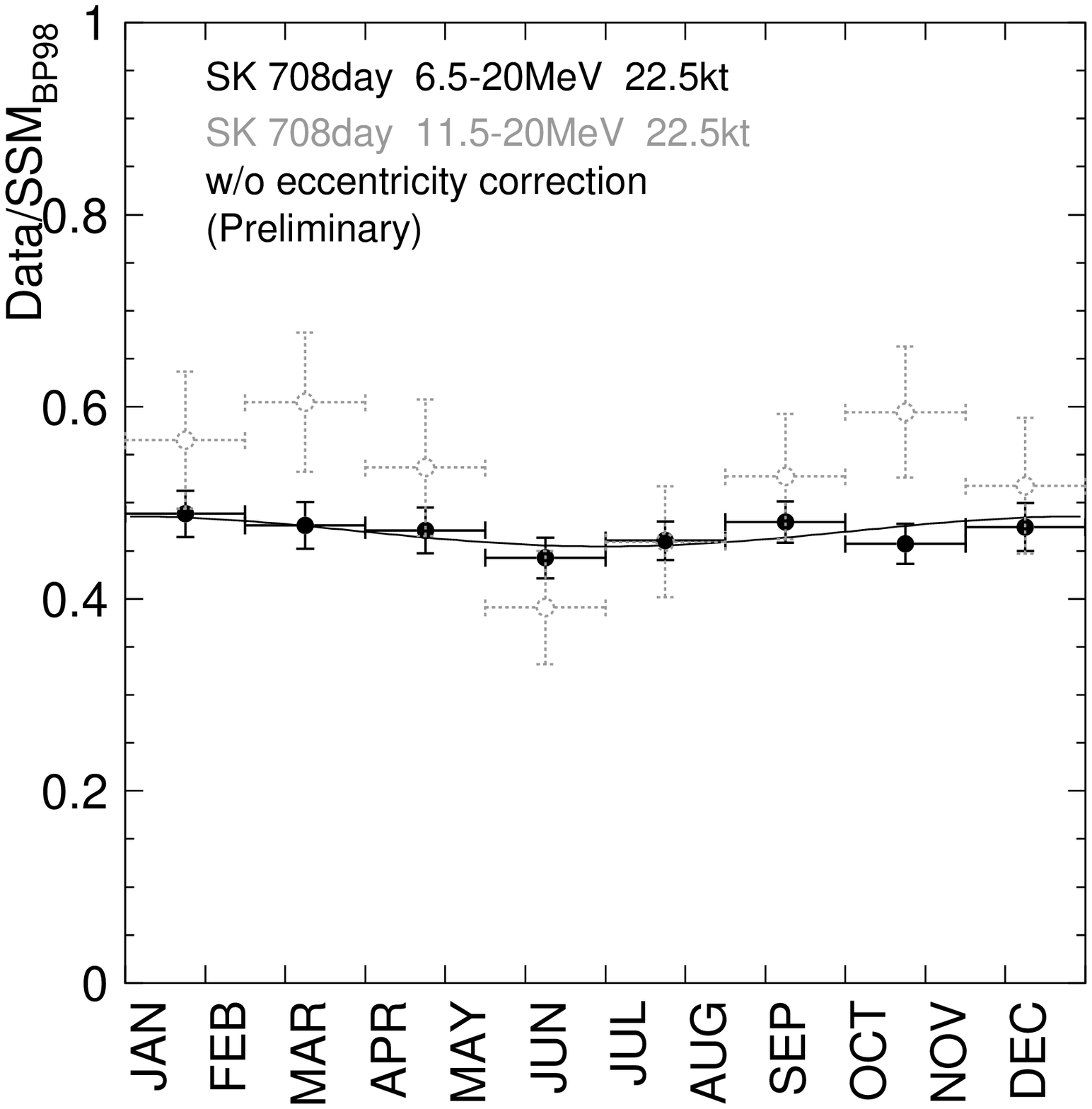}
\epsfxsize 3.5 truein 
\epsfbox{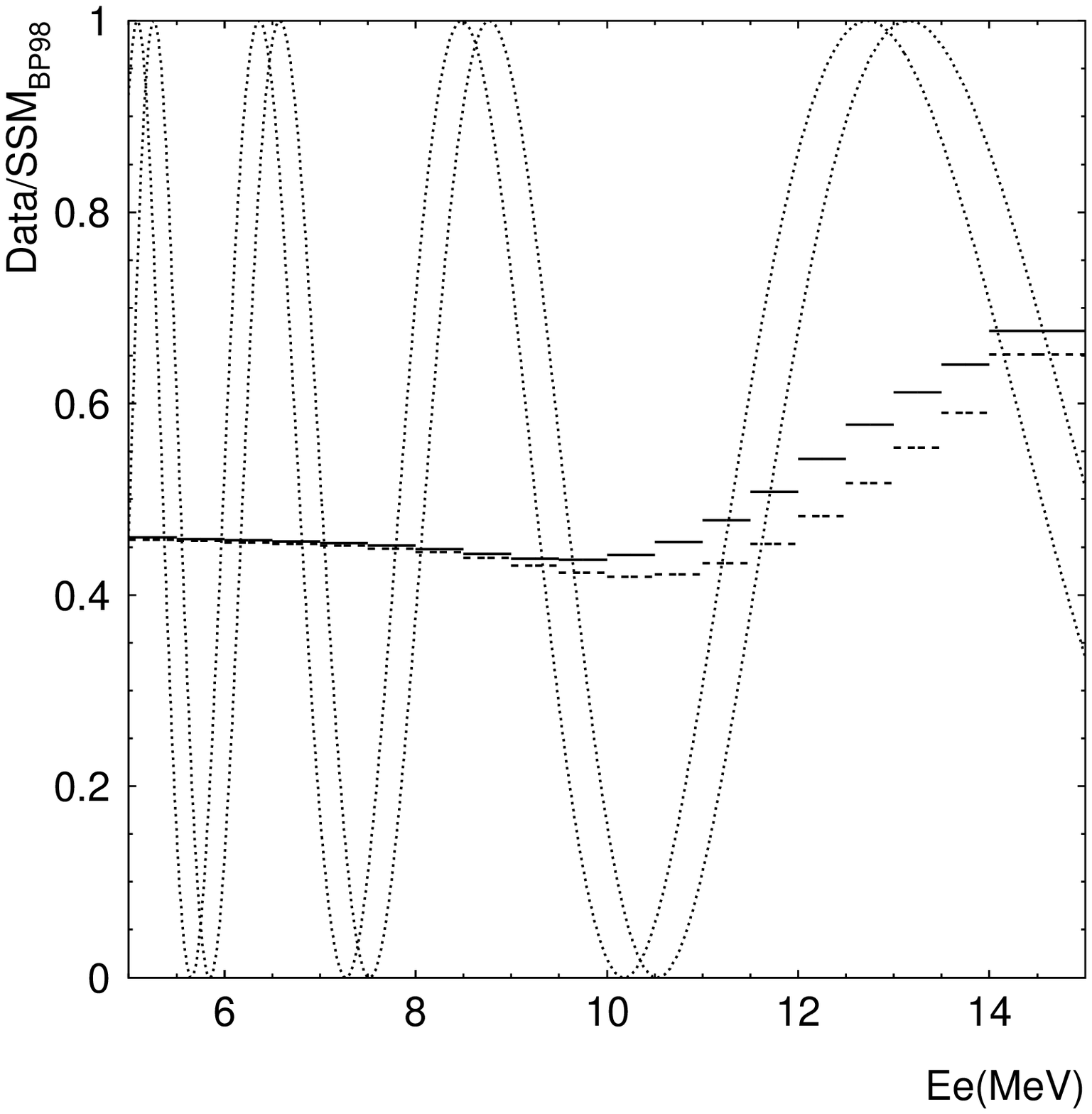}
\vspace*{0.3in}

\caption[]{
\label{seasonal}
\small a) Seasonal Variation \hspace*{1.8in}
b) Expected Spectrum at Perihelion and Aphelion\\
The seasonal variation for the flux above
6.5 MeV (black filled in circles) and 11.5 MeV (grey open circles)
shown in a) is not corrected for the variation
expected from spherical symmetry (black line).
Below 10 MeV the variation in the recoil electron
normalized spectrum is washed out.}
\vspace*{0.2in}
\end{figure}

Since the distance sun --- earth changes with a yearly cycle, vacuum
oscillations can show a seasonal variation of the neutrino flux
(see figure~\ref{seasonal}). Due to the spherical symmetry of the
neutrino flux, this flux should be proportional to $\frac{1}{r^2}$
and show a seasonal variation even without the presence of neutrino
oscillation. The flux above 6.5 MeV recoil electron energy cannot show this
variation, since the low recoil electron energy events dominate the
flux and the variation is washed out for these events
(see figure~\ref{seasonal} b)). After less than three years of
data taking, the statistical accuracy is insufficient to observe this
effect. The data slightly favors the presence of seasonal variation
above 11.5 MeV in addition to the $\frac{1}{r^2}$ variation.
More data is needed to observe a significant effect.

\vspace*{0.5in}

\section{Summary and Conclusion}

After less than three years of data taking, Super-Kamiokande
starts to constrain neutrino flavor oscillation solutions to
the solar neutrino problem. So far, flux normalization independent
signatures are not yet significant (day--night and seasonal
flux variation) or may be explained within the SSM (increase
of the normalized spectrum at the high energy end).
More data and a lower SLE analysis threshold is needed
to measure spectrum and flux variations more
precisely. Due to its high neutrino interaction rate,
Super-Kamiokande is uniquely suited to explore day--night
and seasonal variation which require a large data sample.
Super-Kamiokande is currently still limited by statistics
and could in a few more years perhaps find (or rule out)
neutrino oscillation as an answer to the solar neutrino problem.

\bigskip\bigskip

\centerline{\bf Acknowledgements}
\bigskip\bigskip

We gratefully acknowledge the cooperation of the Kamioka Mining
and Smelting Company. This work was partly supported by the
Japanese Ministry of Education, Science and Culture and the
U.S. Department of Energy.

\end{document}